# Assessing the capacity and flow of ecosystem services in multifunctional landscapes: evidence of a rural-urban gradient in a Mediterranean small island state

Mario V Balzan[*], Julio Caruana, Annrica Zammit

Laboratory of Terrestrial Ecology, Institute of Applied Sciences, Malta College of Arts, Science and Technology (MCAST), Paola, Malta.

[*]Corresponding Author. E-mail: mario.balzan@mcast.edu.mt

# Abstract

Distinguishing between the capacity of ecosystems to generate ecosystem services (ES) and the actual use of these service (ES flow) in ES assessment and mapping is important to develop an understanding of the sustainability of ES use. This study assesses the spatial variation in ES capacity and flow in the Mediterranean small island state of Malta. The services included in this study were crop provisioning, beekeeping and honey production, fodder and livestock production, crop pollination, air quality regulation, and aesthetic ES. This assessment develops different spatial models, which make use of available datasets, causal relationships between datasets, including a generated land use land cover (LULC) map, and statistical models and indicators based on direct measurements. Individual ES indicators were mapped to visualise and compare their spatial patterns across the case study area. Subsequently, an analysis of ES associations and bundles was carried out using Pearson parametric correlation test, for both ES capacity and flow indicators generated from this study, and through Principal Component Analysis. Results demonstrate several significant synergistic interactions between ES capacity and flow in rural landscapes characterised with agricultural and semi-natural LULC categories, indicating high landscape multifunctionality. In contrast, predominantly urban areas tend to be characterised with a low ecosystem capacity and ES flow, suggesting that ES delivery in the landscapes of the study area is determined by land use intensity. These findings support the notion that multifunctional rural landscapes provide multiple ES, making an important contribution to human well-being, and that land use planning that develops green infrastructure in urban areas can significantly contribute to support biodiversity and ES delivery.

# 1. Introduction

The assessment and mapping of ecosystem services (ES) is increasingly recognised as being important to understand the link between ecosystems and their benefits, and their value to human societies for more informed decision-making (Alkemade et al., 2014; Crossman et al., 2013; Jordan and Russel, 2014) and for the development of nature-based solutions as part of sustainable development strategies (Maes and Jacobs, 2017). ES mapping is also a key objective of the EU Biodiversity Strategy to 2020[1], which calls on member states to assess the state of ecosystems and their services in their territory.

This study assesses the spatial variation of ES in a multifunctional Mediterranean landscape, supporting cultural, ecological and economic functions. Two distinctive hallmarks of multifunctionality are that landscapes are considered as a matrix, with high spatial heterogeneity, and as an integrative system defined in terms of ecosystem functions and services (Selman, 2009). Given the diversity of the ecosystems and the interactions between these in an integrative system, multiple ecosystem services are provided by multifunctional landscapes, with manifold services occurring in an inter-related manner within the landscape (O'Farrell and Anderson, 2010) whilst supporting human quality of life (Potschin and Haines-Young, 2016). These benefits depend on the needs, choices and values of people and are also place-related depending on the context (Potschin and Haines-Young, 2013). The maintenance of this multifunctionality, by integrating landscape use in an ecological fabric that maintains ecosystem capacities and flow, is seen as being particularly important in order to achieve sustainability (Lovell and Johnston, 2009; O'Farrell and Anderson, 2010), and may serve as an adaptive strategy to address unknown future conditions through increased resilience (Selman, 2009).

Landscapes within the Mediterranean region have been shaped through natural processes and a long history of human activities, which gave rise to mosaic landscapes characterised by a high diversity of ecosystems (Blondel et al., 2010; Naveh, 1994). These multifunctional landscapes result from a co-evolution of social and ecological systems, and are associated with a high endemicity and species richness but are also of socio-cultural significance (Blondel et al., 2010; Martín-López et al., 2016). In these cultural landscapes, high biodiversity and resilience are particularly linked to the cultural values and to social behaviour and perceptions (Oteros-Rozas et al., 2014).

Multiple ES are provided by the cultural landscapes of the Mediterranean basin, and these contribute to an improved human well-being. However, ES are affected by different direct and indirect drivers of change which alter land uses (Ales et al., 1992; Aretano et al., 2013; Rodríguez-Loinaz et al., 2014; Tzanopoulos and Vogiatzakis, 2011). The intensification of land-use management is associated with the loss of traditionally heterogeneous landscapes, as a consequence of population growth, industrialisation, urban development and increasing tourism. These may threaten the natural capital of the region, as multifunctional landscapes, which have traditionally hosted Mediterranean ecosystems and their services, are lost (Plieninger et al. 2014). These changes can have an even more important effect when they occur in small Mediterranean islands, characterised by a mosaic of land-covers and landscapes (Vogiatzakis et al., 2008) and where the socio-economic and environmental insularity often strengthens the linkages between ecosystems and communities (Balzan et al., 2016). In many Mediterranean islands, traditional human activities, which have shaped the islands' landscapes are almost exclusively related to subsistence production such as mining and agriculture (Petrosillo et al., 2013). These activities have declined in recent decades while tourism has increased substantially and today dominates the local economies of many Mediterranean small islands (Aretano et al., 2013; Petrosillo et al., 2013; Tzanopoulos and Vogiatzakis, 2011). And, whilst traditional rural landscapes, characterised by a mosaic of arable agriculture, and semi-natural ecosystems, have shown a balance between biodiversity and land use in the past, the recent intensification of land use has also led to severe degradation of ecosystems and natural capital (Makhzoumi and Pungetti, 2008).

---

[1] http://ec.europa.eu/environment/nature/biodiversity/strategy/index_en.htm

The aim of this study is to assess how ES capacity and flow vary spatially in the landscapes of Malta (Central Mediterranean). In order to make ES maps operational for landscape planning various have called for a clearer distinction between the different components linking ecosystems to socio-economic systems (Burkhard et al., 2012; Mouchet et al., 2014; Schröter et al., 2014; Villamagna et al., 2013). Maps of the capacity of ecosystems to deliver an ES, and of the flow of the ES, that is the actual ES use, can be a useful tool for planners and policy-makers, as they can allow for the identification of unsustainably used areas (Geijzendorffer et al., 2015), and hence the redirection of ES flows to areas with a higher ES capacity or the planning and development of green infrastructure to improve the capacity of ecosystems to deliver key ES in areas with ES capacity and flow imbalances (Lovell and Taylor, 2013). Green infrastructure has been defined by the EU Strategy on Green Infrastructure as a strategically planned network of natural and semi-natural areas with other environmental features designed and managed to deliver a wide range of ES for human society (EC, 2013).

The landscapes of the study area have been moulded by the geo-climatic conditions and human exploitation acting over several millennia, and are associated with a mosaic of small-scale and interacting ecosystems of high socio-cultural significance. An understanding of the spatial variation of ES capacity and flow in Malta can be used to provide an indication of spatial imbalance in ES capacity and flow, which results in an unsustainable uptake of the ES when ES flow cannot be met by current capacity. The study also investigates interactions amongst ES, occurring when multiple ES respond to the same driver of change or when interactions between the ES alter the provision of another. ES interactions can lead to synergies, that is situations in which both services either increase or decrease, or trade-offs between ES (Bennett et al., 2009; Mouchet et al., 2014).Through the use of data for six ES, we analyse the contribution of different ecosystems to ES bundles, defined as sets of ES that appear together repeatedly (Raudsepp-Hearne et al., 2010). The development of an understanding of the contribution of different land uses to ES delivery, and the overlap between multiple ES in multifunctional landscapes, is relevant to the design of spatial policies that promote sustainable land use (de Groot et al., 2010).

## 2. Methods

### 2.1. Conceptual Framework

The distinction between ES capacity and flow builds on the definition of ecosystem services, which considers these as the contributions that ecosystems make to human well-being. The ES Cascade model develops on this definition and links ecosystems to the human society through a chain of components, namely ecosystem structure and processes, functions, services, benefits and values (Potschin and Haines-Young, 2011). Different indicators have been used to assess and map these different components of the cascade model. In particular, several have distinguished between the ecosystems' capacity and the flow of ES (Figure 1). The ES capacity is here defined as the potential of ecosystems to provide services, while ES flow refers to the actual production of the ES (Villamagna et al., 2013). Based on this definition, ES flow differs from ES demand since the flow measurements focus on the contribution of ecosystems to human well-being whilst the ES demand is the expression of the beneficiaries' preferences for specific ES attributes, such as biophysical characteristics, location and timing of availability, and associated opportunity costs of use. Hence, the ES demand may be larger than the ES flow (Schröter et al., 2014). We further developed this approach to analyse how ES capacity and ES flows vary spatially (Figure 1), and hence allowing for the identification of different spatial patterns of these two components, which can lead to unsustainable ES uptake.

An analysis of the interactions between the considered ecosystem services capacities and flows was carried out, in order to identify ES interactions. These associations between ES may be caused by mechanisms that influence the capacity of ecosystems to provide multiple ecosystem services, hence determining multifunctionality, when several ES rely on the same ecosystem processes or when the management of one service affects another (Mouchet et al., 2014). In contrast, the ES flow occurs at the location where an ES is experienced by people (Villamagna et al., 2013), and hence association between ES flows is influenced by socio-economic factors, such as policies, management practices and ES demand, together with the biophysical processes that determine the ES capacity itself (Queiroz et al., 2015). Given that ES capacity and flows are important components of ES, which are influenced by differing factors, we have investigated interactions at both levels of the ES delivery chain.

## 2.2. Study area

The Maltese archipelago is a group of low-lying small islands situated in the Central Mediterranean Sea at 96 km south of Sicily, almost 300 km east of Tunisia and some 350 km north of the Libyan coast (Appendix A). The archipelago is made up of three inhabited islands (Malta, Gozo and Comino) and several uninhabited islets, with a total land area of 316km$^2$. The landscapes of Malta have been shaped over several millennia by the geo-climatic conditions, and human exploitation, but nonetheless host considerable biodiversity; partly a consequence of the interesting biogeography of the Archipelago (Schembri, 1997).

The Maltese Islands also have a long cultural history and the earliest evidence of settlement dates back to around 7000 BP (Patton, 1996). With agriculture being as old as humankind's remote origins on the archipelago, the landscapes of the Maltese Islands have been highly modified over the millennia. The first settlements were associated with deforestation for agriculture, the introduction of livestock and grazing activities (Schembri, 1997). Today agricultural land cover occupies around 51% of the territory, whilst built-up, industrial and urban areas occupy more than 30% of the Maltese Islands (MEPA, 2010). With a population density of 1,346 persons per km$^2$, the highest in the European Union, and a booming tourism industry the Maltese Islands' biodiversity would be expected to be subject to substantial pressure (NSO, 2014). Within this context, the Maltese Islands make for an interesting model for an analysis of the role of mosaic and multi-functional landscapes in the delivery of ES. This analysis is also relevant to national land use planning and policy-making for ecosystem services and human well-being.

## 2.3. Choice of ES

The selection of the ES investigated in this study allows for an analysis of a range of terrestrial provisioning, regulating and cultural services of relevance to the study area. We followed the CICES (Common International Classification of Ecosystem Services) scheme to categorise the services (Haines-Young and Potschin, 2013). Whilst it was not possible to cover the whole diversity of ES within one study, a range of six key ES were selected for this study (Table 1). These services were identified as being important for the study area in Malta's Fifth Report to the Convention on Biological Diversity (MEPA, 2014), and food provisioning, the maintenance of nursery habitats and species, and cultural ES were identified as being particularly relevant to a national assessment of ES in the case-study area (Mallia and Balzan, 2015). Results from a previous study, during which a questionnaire aimed at exploring how locals value biodiversity within the study area was conducted, indicate that a significantly high value is assigned to the regulation of air quality and the delivery of pollination services together with several cultural services, in particular outdoor recreation and recreation (MEPA, 2014).

The choice of ES also reflects the nature of the study area, which is characterised by an insular environment, heterogeneous landscapes with multiple ecosystem functions and actors, and a strong urban presence and increasing tourism trends (Section 2.2 - Study area). Food provisioning services are a source of livelihood (NSO, 2012), can be vital in terms of the economy and food security in island environments (FAO, 2004) and have been found to be highly valued by island communities (Butler et al., 2014; Kenter et al., 2011). Honey production is an activity of significant scientific and socio-cultural value in the study area, where beekeeping was introduced in historical times and which hosts an endemic and threatened subspecies of the honey bee *Apis mellifera ruttneri* (Sheppard et al., 1997). Together with agricultural food provisioning the coastal and marine environment contribute significantly to food provisioning ES (MEPA, 2014) but, given the focus of this study on the terrestrial environment, and since agriculture on islands is often faced by a number of environmental challenges that might influence the ability of populations to ensure food security, food provisioning from agroecosystems is investigated here. Agricultural systems in island environments are often highly vulnerable to climate change, urban and tourism associated development, limited freshwater resources, the loss of traditional agricultural knowledge and systems, and the introduction of invasive alien species (Wong et al., 2005). Cultural services in the form of increased aesthetic value, recreation and ecotourism are also highly valued by tourists and locals (MEPA, 2014), and contribute directly to the economy. Pollination ES are important for delivering key benefits leading to a marginal increase in crop production of market-based or subsistence crops, fibre and forage products. Air quality regulation is particularly relevant for the study area given that this remains a key environment concern, in particular since the EU limit for a number of pollutants is exceeded in certain areas, with domestic emissions of traffic being identified as the main source (EEA, 2016).

## 2.4. Identification and mapping of ecosystems

The assessment of ES within the study area presents a number of challenges that are mostly associated with the availability of land use and other spatial data at relevant scales, and the scale of the existing spatial data. A land use land cover (LULC) map was created through the use of Sentinel 2 satellite images provided by Copernicus (Drusch et al., 2012), and acquired on 21-08-2016. Sentinel 2 is a multispectral satellite developed by ESA, as part of the Copernicus land monitoring system, which acquires 13 spectral bands with the spatial resolution of up to 10 m (Drusch et al. 2012). The image was converted to reflectance by applying a Dark Object Subtraction (DOS) correction, and using the DOS1 method. Images were then processed and mapped by applying a supervised multispectral classification with the maximum likelihood method. Ground truth areas were used during spectral signature creation and for the evaluation of the accuracy, resulting in a high overall classification accuracy of 89.3% (kappa = 0.88). The final classification consisted of a LULC map with a total of 13 classes (Table 2).

## 2.5. Mapping the Capacity and Flow of ES

The assessment and mapping of ES was performed through the use of the developed LULC map for the study area in combination with other data sets, as explained in more detail in this section.

### 2.5.1. Cultivated Crops

Arable land cover from Sentinel 2 satellite images was used as a proxy for crop cultivation in Malta. Average annual national crop production data for 2014 and 2015, available from the National Statistics Office, was downscaled to the arable land cover classes. Since land cultivated by vegetable crops only makes a fraction of rainfed agricultural land, national census data of 2010 was used as a reference for the area of land cultivated with arable crops. Intensive cropland (permanently irrigated arable crop and greenhouse LULC categories) was subtracted from total land cover dedicated to arable crops to obtain an estimate of lower intensity (extensive) crop production. Crop yield data was then downscaled through a further spatial adjustment for crop cultivation intensity, and a weight of 1.25 was applied to intensive cropland and of 0.66 for the extensive ones (Ivanov et al., 2011). Fruit production data was similarly downscaled to the orchard LULC category.

### 2.5.2. Fodder and Livestock Production

Fodder cultivation is carried out extensively within the study area (MEAIM, 2015), with the fodder being harvested with the stem and let to dry and sold to animal breeders. The total area cultivated with fodder for 2013 (NSO, 2014), was used as an indicator for production. Since fodder, mainly wheat, is cultivated in non-irrigated arable land, total land area cultivated with fodder was considered as a fraction of the total area of rainfed cropland to account for the proportion of the area being used for fodder production.

The assessment of the ES flow was carried out through the use of spatial data pertaining to the location of livestock (cattle, pigs, sheep and goats) production in the Maltese Islands in 2013, available from the National Statistics Office. Density data for each locality was used as an indicator of the ES flow.

### 2.5.3. Honey Production

A preference assessment exercise was carried out with beekeepers to determine the characteristics of ecosystems preferred for beekeeping and honey production. Preference assessment is defined as a direct and consultative method used to demonstrate the social importance of ecosystem services by analysing social motivations, perceptions, knowledge and associated values of ecosystem services demand or use (Santos-Martín et al., 2016). Data presented in this study is based on a workshop carried out with a group of 25 beekeepers, during which the participants were asked to identify how the importance of different biodiversity components (habitats and plants) changed with time. The results presented value the collective preferences of the ES users and provide an indication of the seasonal importance of different land covers and plant species. In order to spatially map the ecosystem capacity to deliver this ES, the data of the beekeepers' plant species preferences was linked to the ecosystem type using expert knowledge and literature (Haslam et al., 1977; Weber and Kendzior, 2006). Subsequently, the cumulative relative frequency for each LULC class was used as an indicator of ES provision. This dataset was preferred over the use of the respondents' habitats preference, since these were often characterised by specific named places (e.g. localities) or landscapes where they place their hives, and which tend to be characterised by multiple habitat types (e.g. garrigue and steppe habitats, arable land, etc.). In contrast, all the respondents identified plant species responsible for ES delivery. This approach adopted here follows that used in spreadsheet models that use lookup-tables based on expert judgement to link ES capacity to species occurrence and land cover (Jacobs et al., 2015; Maes et al., 2016a).

The assessment of the ES flow was carried out through the use of spatial data pertaining to the location of bee hives in the Maltese Islands. The number of registered bee colonies per locality was divided by the area of the locality in order to standardise the data.

### 2.5.4. Pollination

The potential wild bee habitat and the visitation probability, based on the distance from nesting habitats, was used as an indicator of ES capacity (Lonsdorf et al., 2009; Schulp et al., 2014). Bees are considered as the main pollinators in most of temperate and Mediterranean ecosystem (Ollerton et al., 2011). Bee habitats were mapped based on the developed LULC map, for which classes were subdivided into two main types of habitat (Schulp et al., 2014). Based on the classification used by Schulp et al. (2014) and expert opinion, the following two bee habitat types were identified:

- Full Habitat (100%) – grassland/steppe; garrigue; woodland;
- Partial Habitat (50%) – irrigated and rainfed arable land; orchards; vineyards; wetlands.

In contrast to Schulp et al. (2014), and since green urban areas were characterised in terms of the dominant vegetation type, this category was not included in this study. In addition, since woodland habitats are intrinsically small and fragmented habitats within the study area, this category was considered as a full habitat, similar to the 'transitional woodland-shrub' land cover in Schulp et al. (2014). Visitation probability was similarly based on the approach used by Schulp et al. (2014), which calculated visitation probability as an exponential decay function with increasing distance to the habitat. This work is based on results obtained from a meta-analysis on the relationship between landscape structure and pollination success (Ricketts et al., 2008). The visitation probability was then calculated by using Eq. (1):

Visitation Probability = $e^{(beta \times Distance\ to\ habitat)}$

The beta parameter was set to −0.00104, which is the average value obtained by Ricketts et al. (2008) for the decay function (Schulp et al. 2014).

Pollination ES Flow for food production was assessed using a methodology based on Gallai et al. (2009). Crops that benefit from biotic pollination were identified and pollinator dependency values assigned (Gallai et al., 2009; Klein et al., 2007). Using the LULC map, the distribution of the crops was determined by the respective LULC class. Downscaled national vegetable and fruit production data, and total land cover for each crop category from the Agricultural National Census (2010), were used as reference for calculating production dependent on biotic pollination within the relevant LULC categories.

### 2.5.5. Air Quality Regulation

The $NO_2$ dry deposition velocity (capacity) on vegetation was considered as a proxy to assess the ecosystems' *capacity* to remove pollutants from the atmosphere. The method used here follows the work by Pistocchi et al. (2010), which estimates deposition velocity as a linear function of wind speed at 10m height and land cover type.

The $NO_2$ removal flux (flow) was based on the predicted concentration of $NO_2$. A Generalised Linear Mixed Model (GLMM) was used to relate ambient $NO_2$ concentration data to environmental variables. The average $NO_2$ concentration in 134 monitoring sites around the Maltese Islands was calculated for data obtained between 2010 and 2012 during a national sampling program by the Malta Environment and Planning Authority (MEPA). Before fitting the model on the data, co-linearity between variables was controlled using variance inflation factors (VIF) based on a threshold of >10 (Kutner et al., 2004), and a square root transformation was applied on the $NO_2$ concentration variable to ensure a normal distribution, which was tested using a Shapiro-Wilk test of normality.

A beyond-optimal model was then fitted on the explanatory variables from buffers, with a diameter of 1 km, relating to the population density, the area of different road types (Openstreetmap, 2014), built up, industrial and commercial, and airport and port areas (Urban Atlas, European Environmental Agency, 2012), elevation and a categorical variable indicating whether the buffer occupied a coastal area. Elevation data for the study area was extracted from the NASA Shuttle Radar Topography Mission (SRTM) 3 arc-second (approximately 90 meters horizontal data) digital elevation model (DEM) covering all of the study area. In order to avoid pseudoreplication arising from sampling in neighbouring points within the same locality and because of the effect of adjacent road type, these two factors were included in the model as random variables. The most parsimonious model was selected by backward selection and comparison of the Akaike information criterion (AIC) values. $R^2$ values were used to assess the amount of variation explained by the model (Nakagawa and Schielzeth, 2013). The $R^2$ has a "marginal" ($R^2_m$) and a "conditional" ($R^2_c$) component to explain the variance of the fixed effects only ($R^2_m$) or by the entire model consisting of both the random and fixed variables ($R^2_c$). The most parsimonious model was then used to predict the $NO_2$ concentration in a grid with $1km^2$ cells. The predicted point data was then interpolated using inverse distance weighting.

Annual $NO_2$ removal was estimated as the total pollution removal flux in the areas covered by vegetation, extracted from the LULC produced in this study, and was calculated as the product of $NO_2$ concentration and deposition velocity maps (Nowak et al., 2006).

### 2.5.6. Physical use of landscapes

The number of habitats protected in Annex 1 of the EC Habitats Directive (Council Directive 92/43/EEC) was used as a proxy for the capacity of ecosystems to provide opportunities for experiential uses of landscapes. Point values, extracted from 1 $km^2$ grid cells, were interpolated using inverse distance weighting.

The ES flow was measured through the use of a questionnaire carried out with a total of 283 residents. During this study, the respondents were asked to identify 3 places and landscapes that they have visited and are of high aesthetic value, the predominant land use/cover of each site, and the recreational activities they normally carry out at these locations. Frequency data from this preference assessment was then mapped for the identified sites.

### 2.6. Integrated Assessment of Ecosystem Services

An assessment of the spatial variation of ES, and their capacities and flows, was carried out through a statistical analysis of the generated ES maps. Total ES capacity and flow were calculated in an overlaying grid with $1km^2$ cells. The data was then centred and scaled, producing standard Z-scores for each ecosystem service, and checked for multivariate normality. In order to analyse the spatial variation of the ES a Principal Component Analysis (PCA) was carried out. To provide an indication of the association between the spatial overlap of ES and the different LULC classes, cover data for each of these was fitted on the PCA ordination plot using the R Vegan package 'envfit' function (Oksanen et al., 2016). This function fits a centroid of levels of a class variable, and calculates an $R^2$ value as a measure of separation among the different levels of that variable. Additionally, a significance value for the $R^2$ was calculated using 1000 random permutations of the category levels. Correlation analysis was used to identify the existence of synergies and trade-offs, following Mouchet et al. (2014), and using Pearson parametric correlation test both at the ES capacity and flow sides. Similar to Queiroz et al., (2015), the aim of this analysis was not to quantify every specific interaction for each ES pair but to identify weak and strong relationships among multiple ES. The complementary use of correlation analysis and PCA allows for the identification of synergies and trade-offs between ES (Mouchet et al., 2014).

In order to assess the contribution of green infrastructure to the ES capacity and flow within the study area, average Z-scores for each cell in the overlaying 1km$^2$ grid were calculated. A regression analysis was used to estimate the relationship between the ES score and ecosystems contributing to the green infrastructure within each grid. Given that green infrastructure is considered as being multifunctional, providing a wide range of ES (EC, 2013), the cover of ecosystems contributing to the delivery of multiple ES was summed up for each 1km$^2$ cell.

We used QGIS 2.18 Las Palmas Geographical Information Systems to produce ES capacity and flow maps, and all data analyses were conducted in R (R Core Team, 2016).

## 3. Results

Capacity and flow distribution maps of ES are shown in Figure 2 and Figure 3. These results suggest an important contribution of agricultural landscapes in the delivery of the investigated provisioning and regulating ES. On the other hand, habitats of conservation value (capacity) and sites of high aesthetic value (flow) were associated with predominantly coastal landscapes, indicating a potential overlap between capacity and flow. In contrast, agricultural LULC were particularly important for the delivery of provisioning (crop, fodder and honey production) and regulating ES (air quality regulation and crop pollination).

Crop production capacity within the study area was assessed through the use of the Sentinel 2 satellite images by the identification of extensive and intensive plots used for arable crop production (Appendix A). Irrigated cropland during the time of acquisition of the satellite images consisted of relatively small parcels of contiguous fields (1488.7 ± 2846.2 m$^2$). In contrast, arable land that is not permanently irrigated and bare soil cover LULC class (A.2.A, Table 2) occupied 40.82% of the terrestrial land cover. Crop (Figure 2a) and fodder production (Figure 2b) maps indicate that extensive cultivation in non-irrigated arable land contributes significantly to food security, while more intensive permanently irrigated land, associated with higher production per unit area, is composed of small scale agricultural areas. The preference assessment with beekeepers yielded a list of habitats and species that are preferred by beekeepers during the different seasons. Results indicate that the ES capacity of ecosystems varies seasonally, with cultivated agricultural land identified as the most important forage habitat contributing significantly to ES capacity. This changes in summer, when garrigue habitats were preferred by the respondents (Figure 4a). A total of 27 plant taxa were identified as being important for the delivery of this ES. However, the importance of these species varied temporally (Figure 4b). Agricultural land (rainfed crop, orchard and vineyards, and irrigated arable land), grassland and garrigue ecosystems were associated with a relatively high ES capacity, while in contrast built-up areas and sparsely and unvegetated LULC were associated with a low abundance of the identified plant species leading to a low ES capacity (Figure 2d). Whilst ES flow was only analysed at the locality level, these results indicate that most of the beekeeping activities were mostly associated with rural areas (Figure 2e).

Pollinator visitation probability, based on habitat suitability, was used as a measure of pollination ES capacity. Pollinator habitats as defined in this study were associated with agricultural and semi-natural land LULC categories, with full and partial habitat amounting to a total of 15.56% and 51.30% of the study area respectively. Results indicate that rural landscapes, characterised by cultivated and semi-natural LULC categories, are characterised by a higher pollinator visitation probability (Figure 3a). Contrastingly, urban areas have a lower ES capacity but this trend breaks down in the presence of urban green areas LULC categories, which were considered as a suitable habitat for pollinators. Dependency on pollination between different crops varied widely from no dependency to 95% in pumpkins and melons (Table 3). Total average dependence on biotic pollination, for all crops recorded in Klein et al. (2007), was 18.18%. However, this ranged from an average 10.20% in orchards to 19.15% in arable land LULC (Figure 5).

Nitrogen dioxide ($NO_2$) concentration was significantly influenced by the average population density, mean elevation, and the area of trunk, primary and residential roads, and industrial zones (Appendix B). Predicted $NO_2$ values from the most parsimonious model were higher in more urbanised and industrialised environments associated with the eastern and southern zone of the study area. In contrast, the highest $NO_2$ deposition velocity was recorded in woodland habitats and agricultural LULC. ES capacity was the highest in these habitats when located in an urban environment with elevated $NO_2$ concentration.

The frequency by which different places of aesthetic value were mentioned in a questionnaire aimed at investigating the spatial variation in cultural ecosystem services was used as a proxy of ES delivery, while capacity was measured in terms of the distribution of habitats of European Community importance. Coastal sites were associated with a relatively higher ES capacity and flow. Agricultural areas and urban areas were associated with a high ES flow, with the majority of responses in the urban category identifying green urban areas. Participants also identified several outdoor activities carried out within these sites, with walking (20.85%), swimming (15.15%) and picnicking (13.33%) being the most frequent activities associated with these sites (Figure 6).

Our results indicate that ES capacity and flow for cultural ES overlap spatially, with several synergistic interactions being observed between the mapped ES (Figure 7a). Of all the possible pairs of interactions, 47 were significantly correlated and 21 of these pairwise interactions were highly correlated (Pearson coefficient, $r \geq 0.5$) while only one weak negative interaction was recorded (r=-0.13). These results may be explained by an analysis of the spatial overlap of the ES, which was carried out through the use of PCA to analyse spatial overlap of ES capacity and flow (Figure 7b). In general, the PCA results demonstrate how the multifunctional landscapes of the study area, characterised with semi-natural and agricultural habitats, are associated with the delivery of multiple ES. Principal component 1 (PC1) corresponded to an axis that varied from urban to agricultural land, and explained a total of 51.7% of the total variance of ES data. All ES were positively related to PC1. Principal Component 2 (PC2) explained 12.9% the ES variance, and corresponded to a gradient from natural habitats (woodland, grassland and garrigue) to agricultural land uses (Appendix C). The remaining principal components explained less than 10% each of additional variance in services. These results indicate a relatively strong association between garrigue, steppe and woodland ecosystems with aesthetic (cultural) ES capacity and flow. Predominantly agricultural LULC were associated with provisioning and regulating ES, namely crop, fodder, honey production, and crop pollination and air quality regulation). Ecosystems considered as forming part of the terrestrial green infrastructure of the study area were associated with a high ES capacity and flow (Figure 8). Urban areas were associated with low capacity and flow of ecosystem services, with the exception of livestock provisioning which appears to be associated with relatively low density urban areas characterised with landscapes with both urban and agricultural land cover. However, in some cases capacity and flow vary with the LULC and hence also spatially. As an example, $NO_2$ deposition velocity was higher in predominantly agricultural landscapes whilst $NO_2$ removal flux was highest in green infrastructure located in an urban setting (Figure 3).

## 4. Discussion

4.1. Spatial variation and synergies between ecosystem services

We have mapped multiple ecosystem services in a Mediterranean island system, by using both readily accessible datasets and empirical data, and this has revealed substantial variation in ES capacity and flow. Results provide evidence of the link between ecosystems and the ES in the study area and demonstrate how rural landscapes, characterised by patches of semi-natural and agricultural areas, are important for the delivery of multiple ES.

The ES capacity is associated with semi-natural LULC categories (woodland, garrigue, grasslands) for aesthetic (cultural) ES and with agricultural LULC categories contributing significantly to provisioning and regulating ES. This general pattern is a consequence of the type of land use, with certain ecosystems being more effective at increasing the ES capacity, and the dominance of agricultural LULC within the landscapes of the study area. An example of this can be seen with air quality regulation, where ES capacity ($NO_2$ deposition velocity) is highest per unit area in the woodland LULC category making up a small fraction of the total case-study area whilst the highest ES flow per unit area ($NO_2$ removal flux) is recorded in woodland areas located in an urban environment associated with elevated $NO_2$ concentration. In contrast, agricultural LULC categories provide an overall higher ES capacity and flow due to a more extensive land cover. These patterns complement observations made by Baró et al. (2016), who found the highest capacity of air quality regulation and outdoor recreation ES in natural areas located on the outskirts of the Barcelona metropolitan region but the highest flow in peri-urban and suburban green areas. Similarly, a general decline in ES with increasing distance from protected areas was recorded in Germany and Poland (Łowicki and Walz, 2015). Within this study, food provisioning ES and crop pollinator dependence were mainly associated with agricultural land cover classes. However, actual use (i.e. the flow) of provisioning ES in some cases occurs in other LULC categories, mainly in low density urban areas in the case of livestock production and in landscapes characterised with semi-natural habitats in the case of honey production. A distinction between the capacity and flow in this case allows for the identification of a potential spatial imbalance between the habitats identified as being important for beekeeping and honey production and the actual use of the service, measured by the location of beehives.

The results provide an indication of the multifunctional nature of the agricultural landscapes of the study area as these are characterised by various functions responsible for the delivery multiple ecosystem services, such as crop, livestock and honey provisioning, pollination and air quality regulation ES, which lead to improved human well-being. Hence, the replacement of these landscapes, characterised by a mosaic of agricultural and semi-natural habitats, would be expected to lead to the reduction of the ES capacity and flow. These observations contrast with those in other studies showing a decline in ES delivery in agricultural landscapes. Intensive agricultural municipalities, characterised by high landscape homogeneity, were shown to provide food products but are relatively poor in terms of capacity to deliver other ES (Baró et al., 2017). Similarly, significant trade-offs have been recorded between provisioning and regulating ES (Raudsepp-Hearne et al., 2010), suggesting that intensive management for maximising production from a provisioning ES may undermine the sustainability of the provisioning ES itself and diminish the possibility of diversifying economic activities (Foley et al., 2005; Raudsepp-Hearne et al., 2010; Rodríguez et al., 2006). Trade-offs recorded in this study were associated with animal husbandry in landscapes characterised by a higher urban land cover. Hence trade-offs were recorded with habitats associated with semi-natural land cover. Livestock production was not strongly associated with the other bundles of ES. Similar results were obtained by Baró et al. (2017), who argue that this may be a consequence of the nature of this ES, which unlike crop production, may not require large parcels of land.

Ecosystems in the multifunctional landscapes of the study area are associated with a number of strong synergistic interactions between ES, providing an indication of the multifunctionality of rural landscapes within the study area as these contribute significantly to the delivery of multiple ES. This supports the notion that diverse and heterogeneous landscapes, resulting from socio-ecological evolution, in the Mediterranean region are tightly linked to the capacity and flow of ES (Martín-López et al., 2016).

4.2. Improving ES delivery across a rural-urban gradient in multifunctionality

Urbanisation in several Mediterranean small islands, is a consequence of contemporary pressures arising from tourism and land commercialisation which, has led to a dramatic changes in cultural landscapes from rural to urban (Papayannis and Sorotou, 2008), with anthropogenic processes operating to increase land-use intensity inside the urban zones (Tzanopoulos and Vogiatzakis, 2011). The results obtained here indicate a general decline in ES capacity along a rural-urban gradient in multifunctionality. Most of the use of the ES, and therefore the flow, also occurs in predominantly rural areas, with the exception of livestock production. This gradient in ES delivery is determined by the land use intensity and the inclusion of semi-natural land cover in urban settings contributes to a higher capacity (Baró et al., 2017; Kroll et al., 2012). Results obtained here are coherent with others indicating a low ES capacity in urban areas, despite these being normally characterised by a high ES demand (Baró et al., 2017, 2016). The lack of adequate green infrastructure in these urban settings leads to a loss of the ES traditionally provided by rural landscapes.

Studies assessing the spatial variation of ES across landscapes are important for the setting up and enhancement of tools to evaluate and integrate ecosystem services in landscape planning processes, as policy-makers can use this information to design spatial policies to optimise the capacity of ecosystems to provide goods and services (de Groot et al., 2010; Roces-Díaz et al., 2014). Urban areas do not necessarily provide fewer ES compared to other regions, as green infrastructure such as the presence of tree cover, can significantly contribute to support biodiversity and ES delivery (Larondelle and Haase, 2013). Indeed, a key observation emerging from this study is that the development of green infrastructure within urban areas can enhance the capacity and actual flow of ES (Elmqvist et al., 2016; Maes et al., 2016b). Other studies have documented how urban green infrastructure may lead to improved availability of flowering plans for pollinators (Hicks et al., 2016), removal of air pollution (Nowak et al., 2006), increased water infiltration and local climate regulation (Pataki et al., 2011), and providing alternative sites for aesthetic, recreation and other cultural ES (Casado-Arzuaga et al., 2013). Semi-natural land cover, such as woodland, shrub and other ecosystems, would be expected to lead to an improved contribution to pollination and air quality regulation and cultural ES. An increase in the size of forest land cover in the landscapes of the Mediterranean Island of Vulcano was associated with an improved ES flow (Aretano et al., 2013) while sustainable urban greening in Mediterranean islands has also been suggested as a conservation strategy to improve the quality of urban living and balance this with landscape protection and biodiversity conservation in Mediterranean islands (Makhzoumi and Pungetti, 2008). Green space managed for food production may also prove catalytic for the enhancement of ES, particularly in urban areas, as human use of these sites may lead to improved vegetation cover, higher plant species richness and improved cumulative ES delivery (Dennis and James, 2016).

4.3. Limitations and Applicability of results

The target of a study should be a good compromise among precision, broad applicability to a variety of landscapes and adaptability to varying data availability (Kroll et al., 2012). Through the adoption of a general approach, which provides comparable and standardised results of relevance to policy-making, this study links ES spatial data to LULC units and analyses the interactions between ES. This assessment is considered as being reliable, given that several proxies used here have been implemented in previous studies (Baró et al., 2016; Jacobs et al., 2015; Maes et al., 2016b; Schulp et al., 2014). Results obtained here offer a number of salient points for landscape management for ES delivery within the study area. Nonetheless, the methodology is challenged by a number of limitations that may lead to uncertainty in the application of the results (see also 2.4 - Identification and mapping of ecosystems).

Limitations were mostly associated with the availability of data at the relevant spatial scale given the very local nature of this case-study and hence the need for finer resolution data. Whilst the obtained mapping data precision is relatively accurate, similarities between LULC categories may influence the precision of the assessment. For example, the similarity of soil cover of fallow land to that in semi-natural communities and that between different types of orchards and other mixed shrub communities compromises the precision of the proxies used here. However, the contribution of similar LULC categories would be expected to offer a smaller variation in ES delivery for most ES. In addition, the use of proxies, based on downscaled national data, for mapping provisioning ES is less likely to capture finer changes in productivity associated with the biophysical environment and management intensity. These indicators can be improved through data collection at a local scale and the incorporation of this in spatial models. Similarly, the role of small and discrete habitats harbouring biodiversity, and often of significant ecological importance such as coastal sand dunes or intermittent streams, may be underestimated in this study while marine ecosystems are also likely to provide important ES within the study area.

Limitations associated with the data availability are also a consequence of the small size of the study area, and are congruent with observations made elsewhere that there is a lack of ES information at local scales that is relevant for decision-making (Burkhard et al., 2012; Hauck et al., 2013; Rodríguez-Loinaz et al., 2014). On the other hand, the size of the study area provides an ideal setting for assessing spatial variation in ES, and their capacity and flow, since synergies and trade-offs between ES tend to be produced at local level rather than at larger scales, and interactions observed at larger scales may not reflect specific trade-offs observed at a local scale (Hauck et al., 2013; Holt et al., 2015). The applicability of the study to analyse spatial capacity-flow balance would also benefit from further refinement of the used ES indicators so that these can be measured in similar units, which could then be subtracted (Schröter et al., 2014).

This study uses proxies to develop an understanding of the spatial variation of ES in a Mediterranean island landscapes but an assessment of the temporal variation of the components of ES delivery remains particularly important. In Mediterranean island landscapes temporal changes in ecosystems occur as a consequence of key pressures affecting these, and which include population and tourism growth and urbanisation, as has been demonstrated elsewhere (Aretano et al., 2013; Petanidou et al.,

2008; Tzanopoulos and Vogiatzakis, 2011). An improved understanding of the ES capacity and flow within these ecosystems provides opportunities for landscape management for a sustainable ES delivery. The importance of finer temporal variations is demonstrated for beekeeping/honey production and cultural ES, in which ES demand and consequently also flow varies with seasons. While this study provides an understanding of the services provided by different ecosystems an improved understanding of the temporal variation in ES capacity and flow would permit for the development of strategies and plans that consider this additional level of complexity for sustainable ES delivery.

## 5. Conclusions

This paper has presented an analysis of the spatial variation of ES in a local-scale study, and contributes to a better understanding of how ES capacity and flow are distributed in a multifunctional landscape. Furthermore, this study provides evidence that agricultural and semi-natural ecosystems provide a range of ecosystem services (capacity) and are also where most of the use of the ES occurs (flow). The present study clearly demonstrates the presence of a rural-urban gradient in multifunctionality, and thus adds to the understanding of the impacts of land use change on ecosystems and their services in the cultural landscapes of a Mediterranean small island state. Hence, these observations confirm the significance of rural landscapes for the delivery of ES, and provide evidence of the need for the development of green infrastructure in urban zones to improve ES capacity and flow. The method advanced in this study can be extended, and used with data for different ecosystem services, and used to develop and implement spatial policies that aim to achieve sustainability of ES capacity and flow.

## 6. Acknowledgements

MB designed the study, analysed the data, wrote the manuscript and supervised work in this study. MB, JC and AZ collected the data, and contributed to the interpretation of the results. MB has received funding from the European Union's Horizon 2020 project ESMERALDA under grant agreement No 642007. We thank two anonymous reviewers whose in-depth comments have improved the paper.

# 8. Tables

*Table 1 – Mapping ecosystem services capacity and flow. The Common International Classification of Ecosystem Services (CICES) category and the respective indicator for ES capacity and flow are shown.*

|   | Ecosystem Service (CICES 4.3) | Indicator | Capacity/Flow |
|---|---|---|---|
| 1 | Cultivated crops | Downscaled crop production (ton/Km$^2$) | Capacity/Flow |
| 2 | Reared animals and their outputs | Beekeepers' Habitat Preference (Frequency of responses) | Capacity |
|   | Reared animals and their outputs | Number of hives/Km$^2$ | Flow |
| 3 | Materials from plants, algae and animals for agricultural use | Rainfed agricultural land (Fodder production potential) | Capacity |
|   | Materials from plants, algae and animals for agricultural use | Livestock (number of Cattle, Sheep, Goats)/Km$^2$ | Flow |
| 4 | Pollination and seed dispersal | Pollinator visitation probability | Capacity |
|   | Pollination and seed dispersal | Crop pollinator dependency | Flow |
| 5 | Dilution by atmosphere, freshwater and marine ecosystems | NO$_2$ deposition velocity (mm/s) | Capacity |
|   | Dilution by atmosphere, freshwater and marine ecosystems | NO$_2$ removal flux (ton/ha/year) | Flow |
| 6 | Physical use of land- /seascapes in different environmental settings | Number of habitats of community importance | Capacity |
|   | Aesthetic | Preference Assessment with locals (Frequency of responses) | Flow |

*Table 2 – Terrestrial ecosystems identified within the land use land cover map. The European Commission Mapping and Assessment of Ecosystems and their Services (MAES) initiative typology of ecosystems is used as a reference and is adapted to the local land uses and cover in this study* (Maes et al., 2013).

| MAES (Maes et al. 2013) code | Terrestrial Ecosystems in MAES | Adapted LULC Code | Terrestrial Ecosystems in LULC map for Malta | Description |
|---|---|---|---|---|
| A.1 | Urban ecosystems | A.1.A | Urban areas | This class includes urban, industrial, commercial, and transport areas. |
| | | A.1.B | Roads | This class includes main road networks within the study area. |
| A.2 | Cropland | A.2.A | Non-irrigated arable land and bare soil cover | Ploughed land, cultivated with cereals, legumes, fodder crops, root crops and fallow, but which has no productive vegetal cover on the date of acquisition. |
| | | A.2.B | Irrigated arable land | Arable land cultivated with crops and having an artificial water supply. |
| | | A.2.C | Orchard and shrub communities | Parcels of land dominated by fruit, berry, citrus and olive plantations, and including agroforestry. |
| | | A.2.D | Vineyards | Areas planted with vines. |
| | | A.2.E | Greenhouses | Greenhouses used for crop production. |
| | | A.2.F | Golf course | Area cultivated with grass and trees, and used for recreational purposes. Considered separately from irrigated arable land cover (A.2.B) when assessing provisioning ES. |
| A.3 | Grassland | A.3 | Steppe communities | Areas dominated by grassy vegetation. |
| A.4 | Woodland and forest areas | A.4 | Woodland | Areas dominated by a mixed woody vegetation of various age. |
| A.5 | Heathland and shrub areas | A.5 | Sclerophyllous vegetation | Areas dominated by bushy sclerophyllous vegetation, including maquis and garrigue. |
| A.6 | Sparsely or unvegetated land | A.6 | Sparsely or un-vegetated rock cover | All unvegetated or sparsely vegetated habitats. These include bare rocks and sandy beaches. |
| A.7 | Inland wetlands | A.7 | Wetlands | Inland and coastal wetlands supporting freshwater and saline marshland communities. |

*Table 3 - Dependency of crops on insect pollination (%) for dependent crop types for total crop production data of 2015, reported by the Malta National Statistics Office (NSO).*

| Crop Type | Dependency (%) | Production dependent on biotic pollination (Kg) | Economic Value of biotic pollination (€) |
|---|---|---|---|
| Grapefruit | 5 | 464.25 | 92.21 |
| French beans | 5 | 2,833.14 | 4,992.02 |
| Tangerines | 5 | 5,524.61 | 4,797.19 |
| Lemons | 5 | 21,420.38 | 15,155.71 |
| Bell pepper | 5 | 43,402.14 | 37,338.63 |
| Oranges | 5 | 59,386.70 | 43,750.44 |
| Tomatoes | 5 | 598,865.88 | 469,551.11 |
| Figs | 25 | 30,427.15 | 58,027.14 |
| Eggplants (aubergines) | 25 | 184,857.20 | 104,242.45 |
| Strawberries | 25 | 197,010.15 | 472,927.95 |
| Broad beans | 25 | 612,982.07 | 205,037.48 |
| Berries (Mulberries and Blackberries) | 65 | 2,077.27 | 5,214.11 |
| Apples | 65 | 18,339.75 | 13,880.43 |
| Apricots | 65 | 24,671.11 | 32,248.80 |
| Plums | 65 | 60,790.32 | 58,476.29 |
| Pears | 65 | 111,133.84 | 123,796.21 |
| Peaches and nectarines | 65 | 497,056.33 | 524,343.37 |
| Cucumbers | 65 | 534,445.43 | 212,055.57 |
| Melons | 95 | 2,834,969.84 | 377,497.72 |
| Watermelons | 95 | 3,454,084.26 | 1,209,996.27 |
| Pumpkins, marrows and gourds | 95 | 4,334,570.25 | 1,128,609.79 |
| | Total | 13,629,312.07 Kg | 8,186,331.78 € |

9. Figures

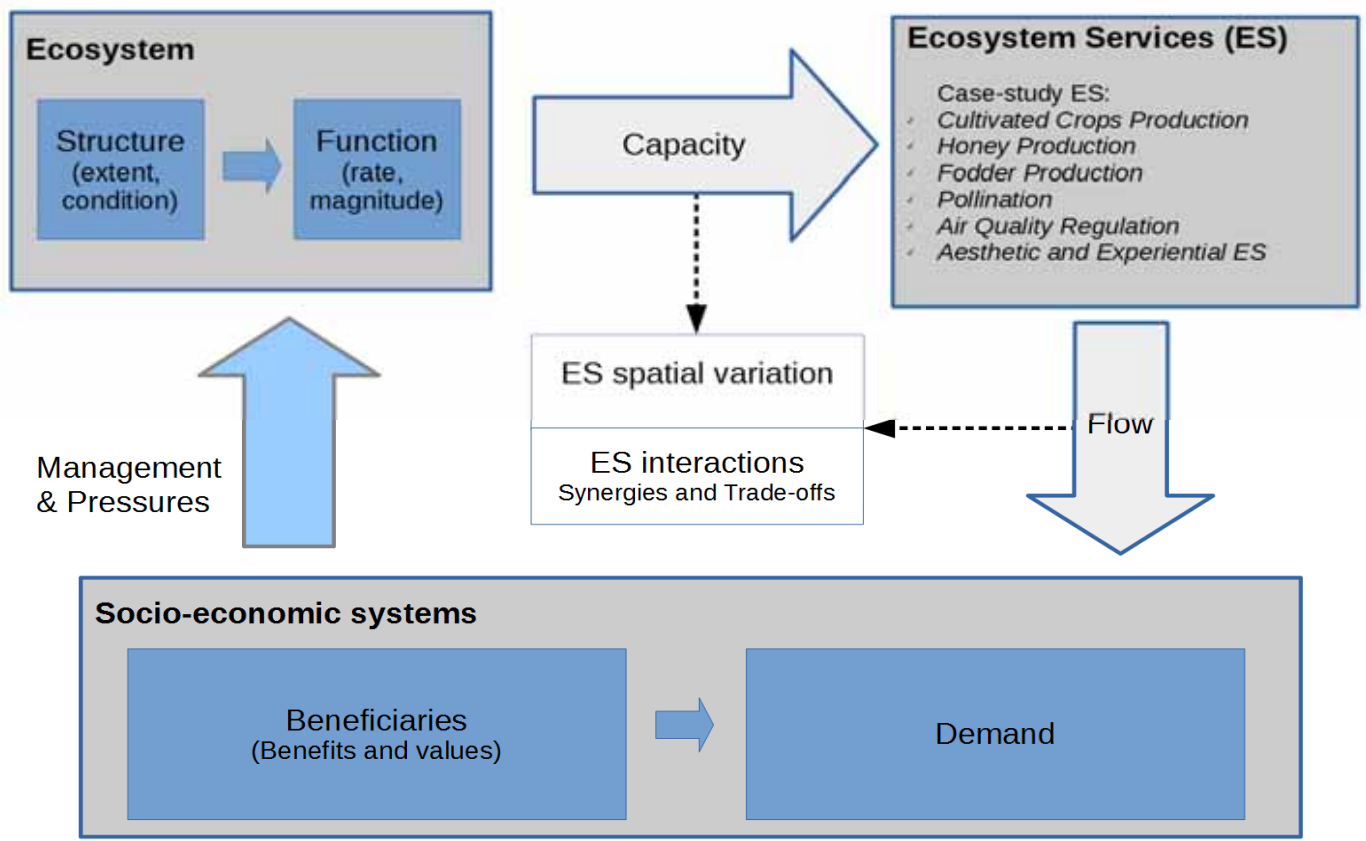

*Figure 1 – General conceptual diagram linking ecosystems' capacity and the flow of ES to human well-being. Block arrows indicate the relationship between the ecosystem, ecosystem services and socio-economic systems, while dashed arrows indicate the level of analysis in this study through the identification of interactions, overlap and synergies and trade-offs in ES delivery.*

a)

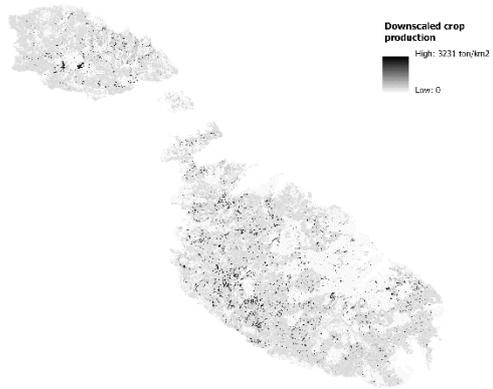

b)

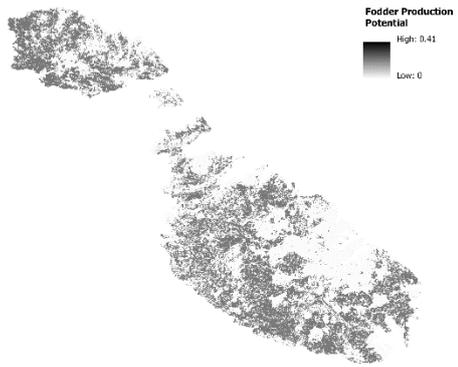

c)

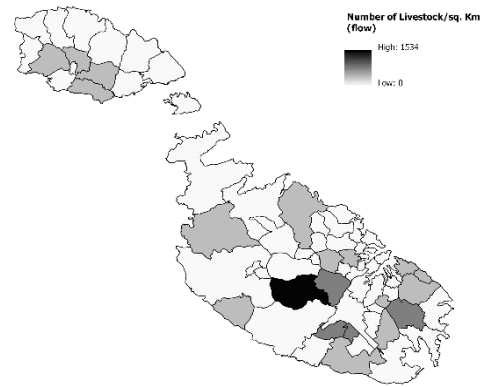

d)

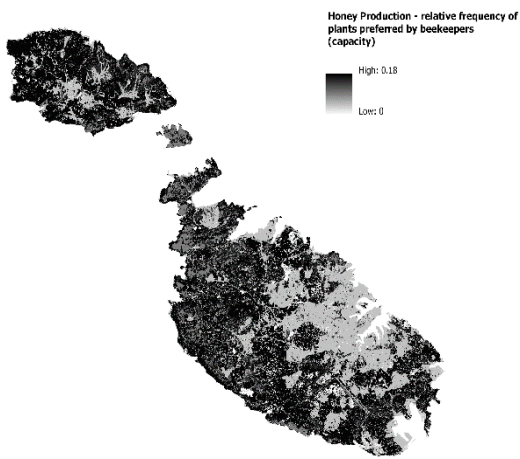

e)

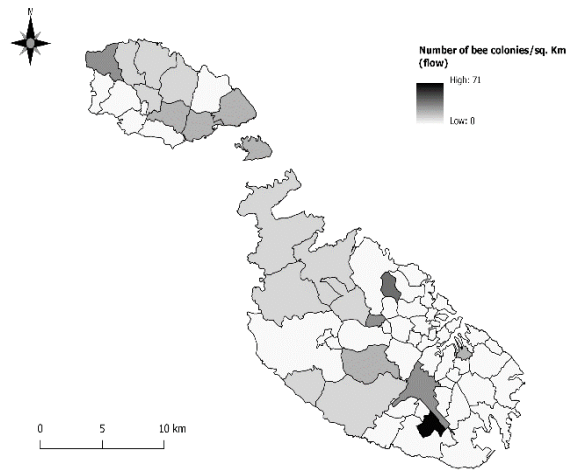

*Figure 2 – Spatial models for provisioning ES capacity and flow.*

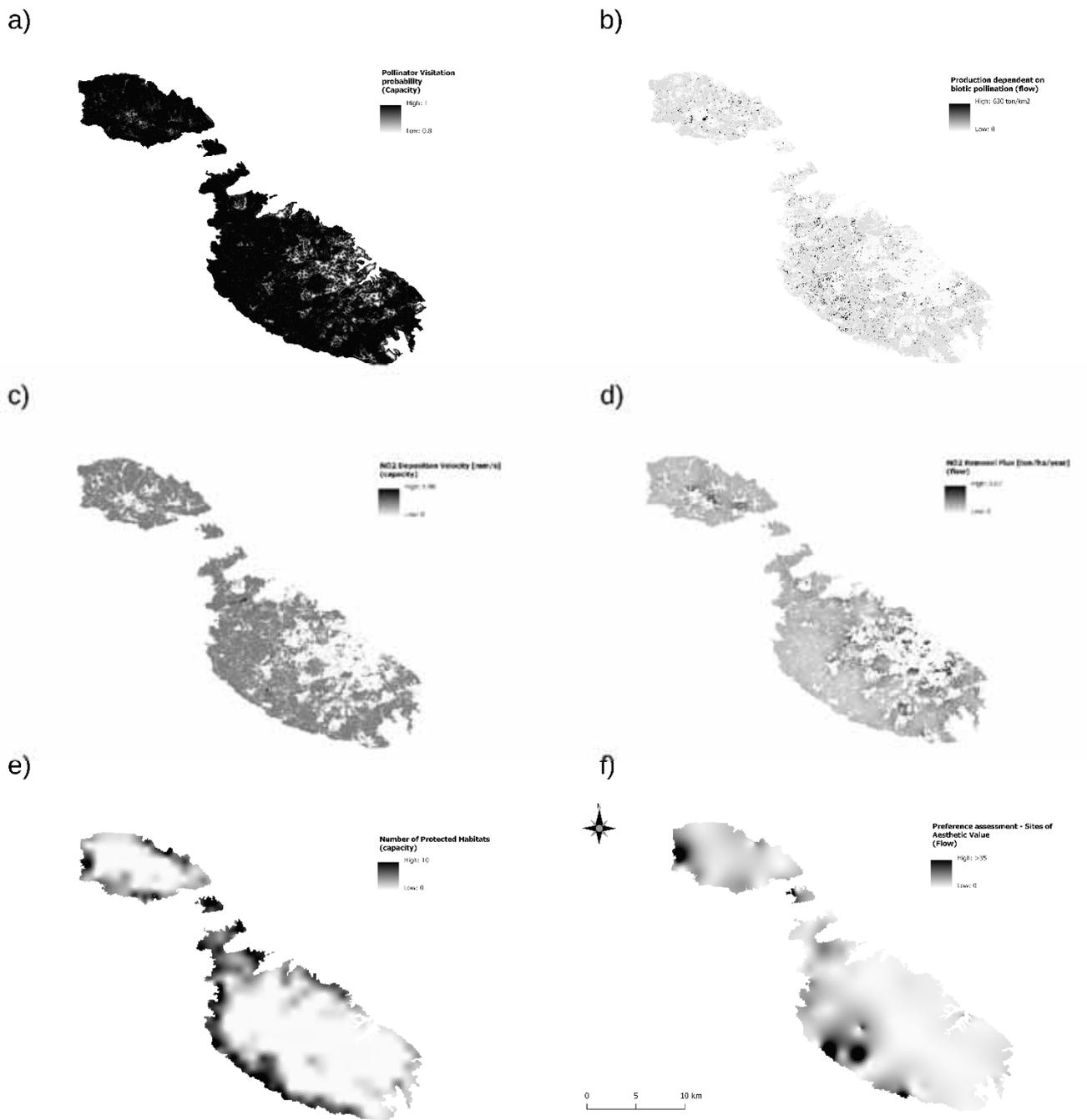

*Figure 3 - Spatial models for regulating and cultural ES capacity and flow.*

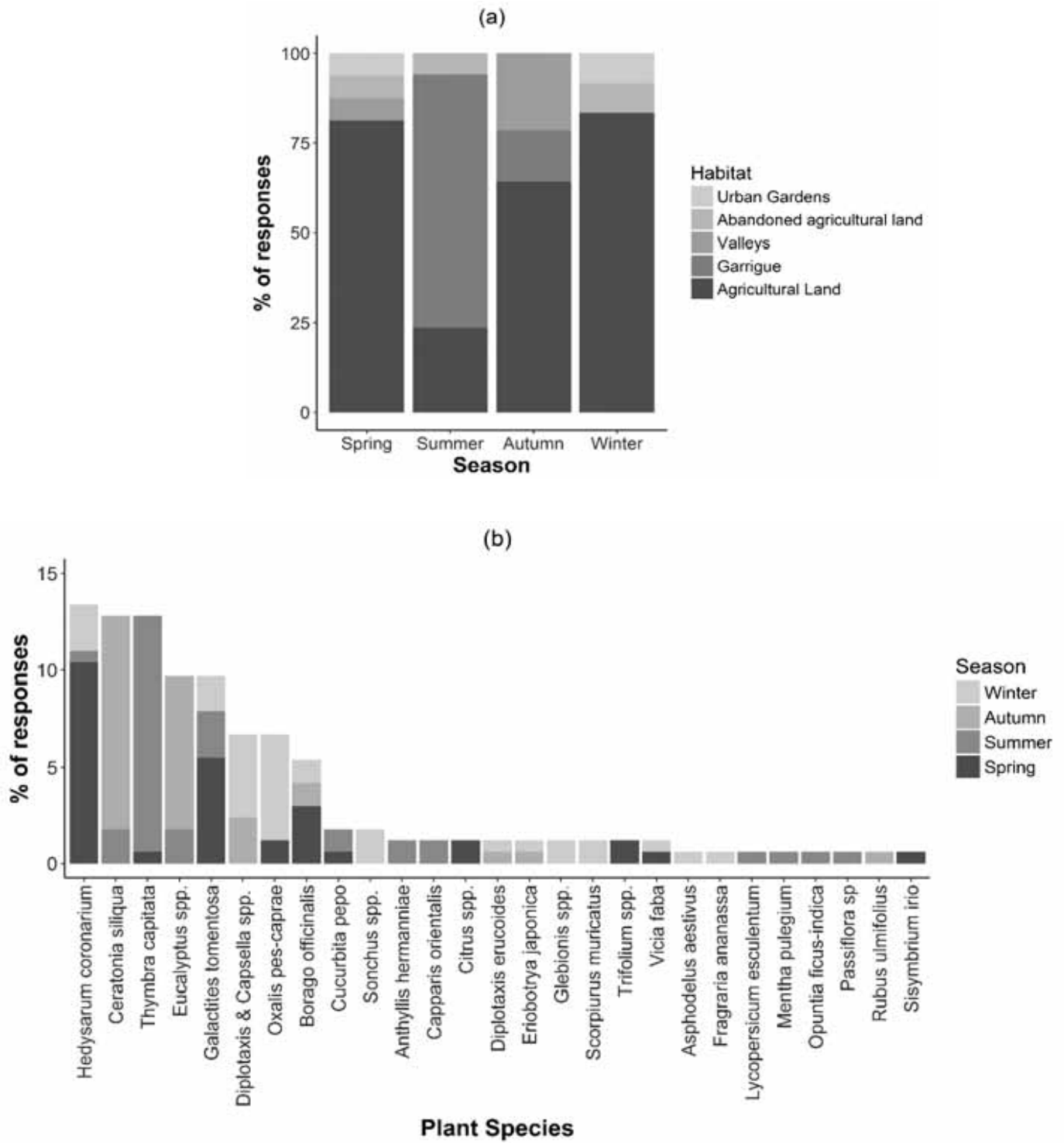

*Figure 4 – (a) Habitats and (b) plant species preferred by beekeepers according to seasons.*

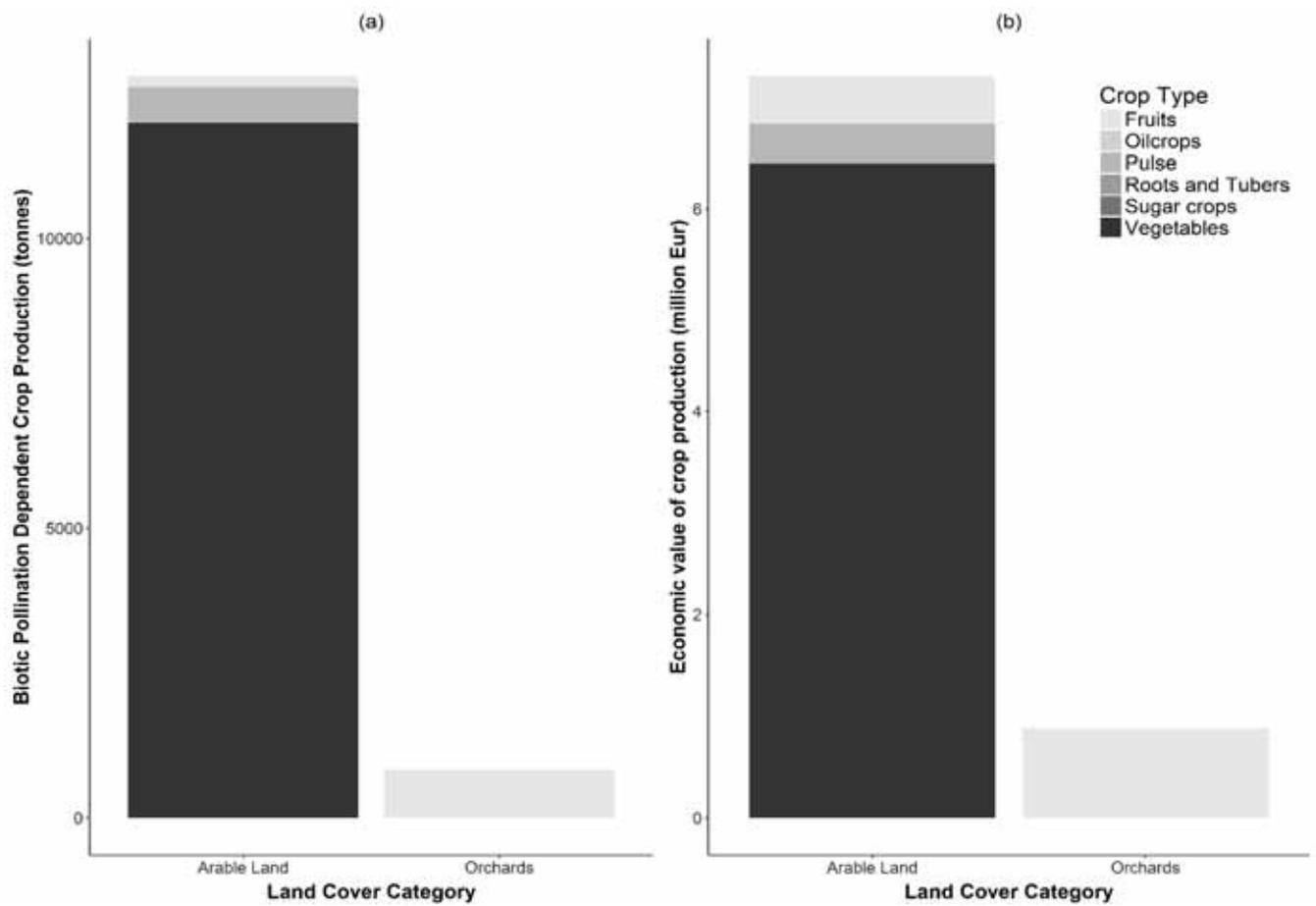

*Figure 5 – (a) Crop production in tonnes dependent on biotic pollination and (b) associated economic value for arable land and land cultivated with permanent crops.*

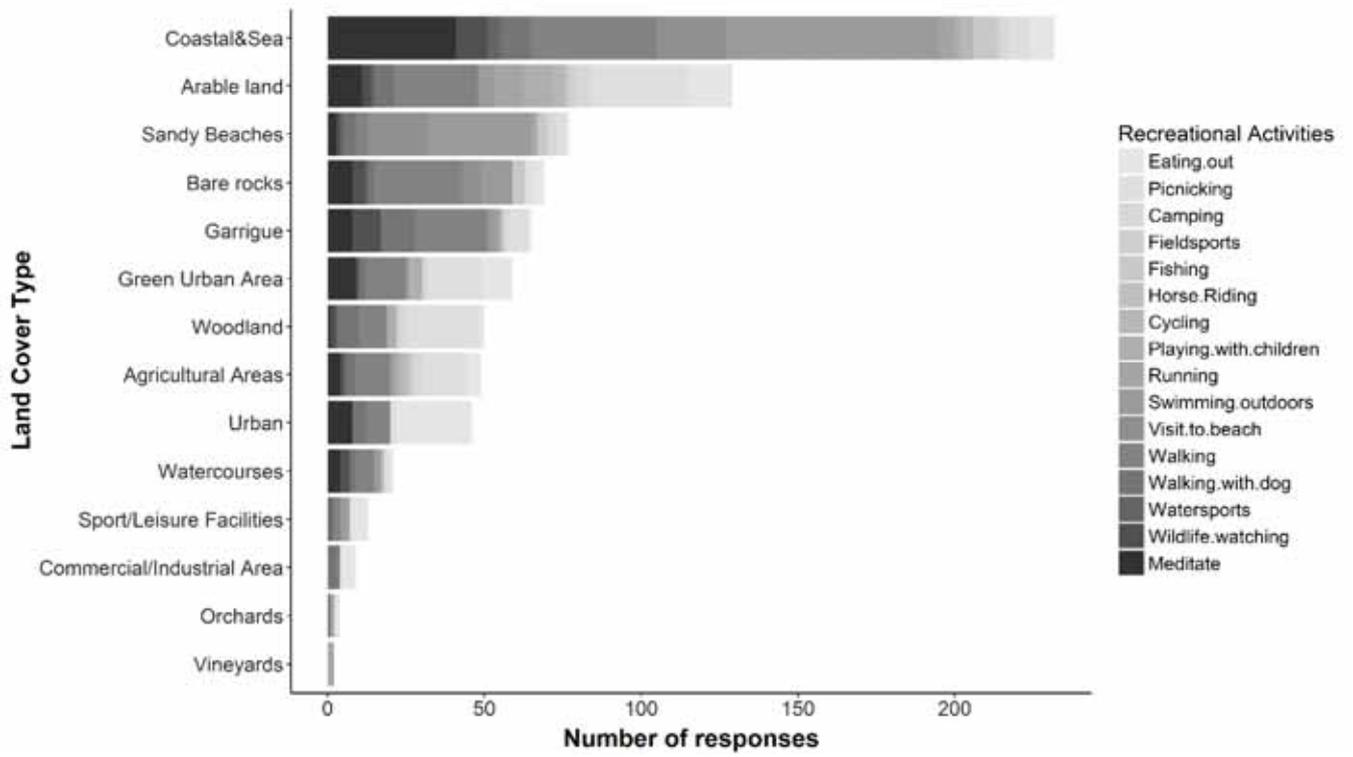

*Figure 6 - The association of the land cover type identified by respondents to landscape aesthetic value (total number of responses) and use for recreational activities.*

(a)

(b)

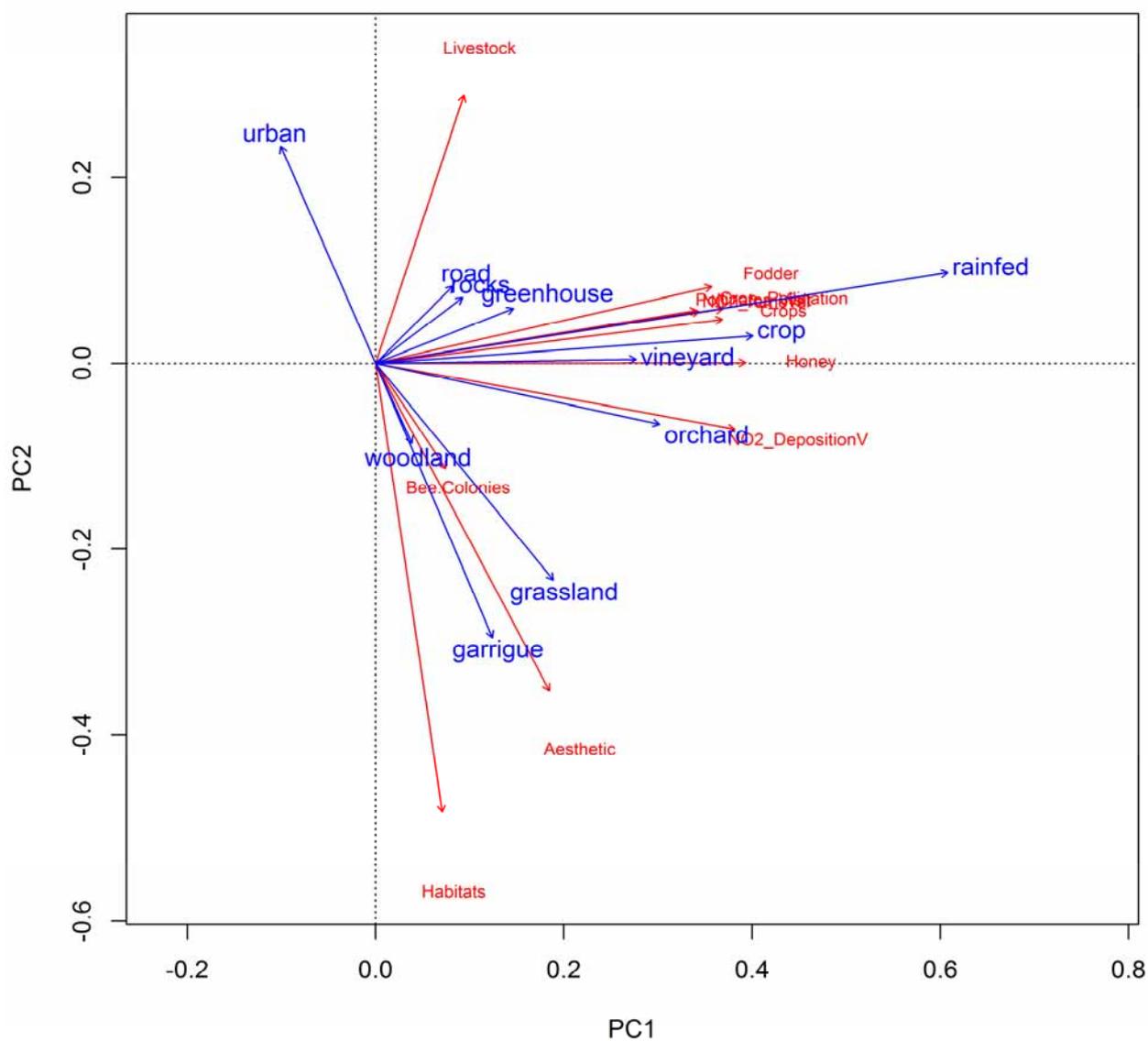

*Figure 7 – (a) Pearson's correlations. Correlation analysis of the pair-wise interactions between ecosystem services (C: ES capacity; F: ES Flow). (b) Principal Component Analysis of multivariate data for the total ES capacity and flow in 1km² grid, with environmental data for LULC category area fitted on the PCA ordination plot.*

(a)

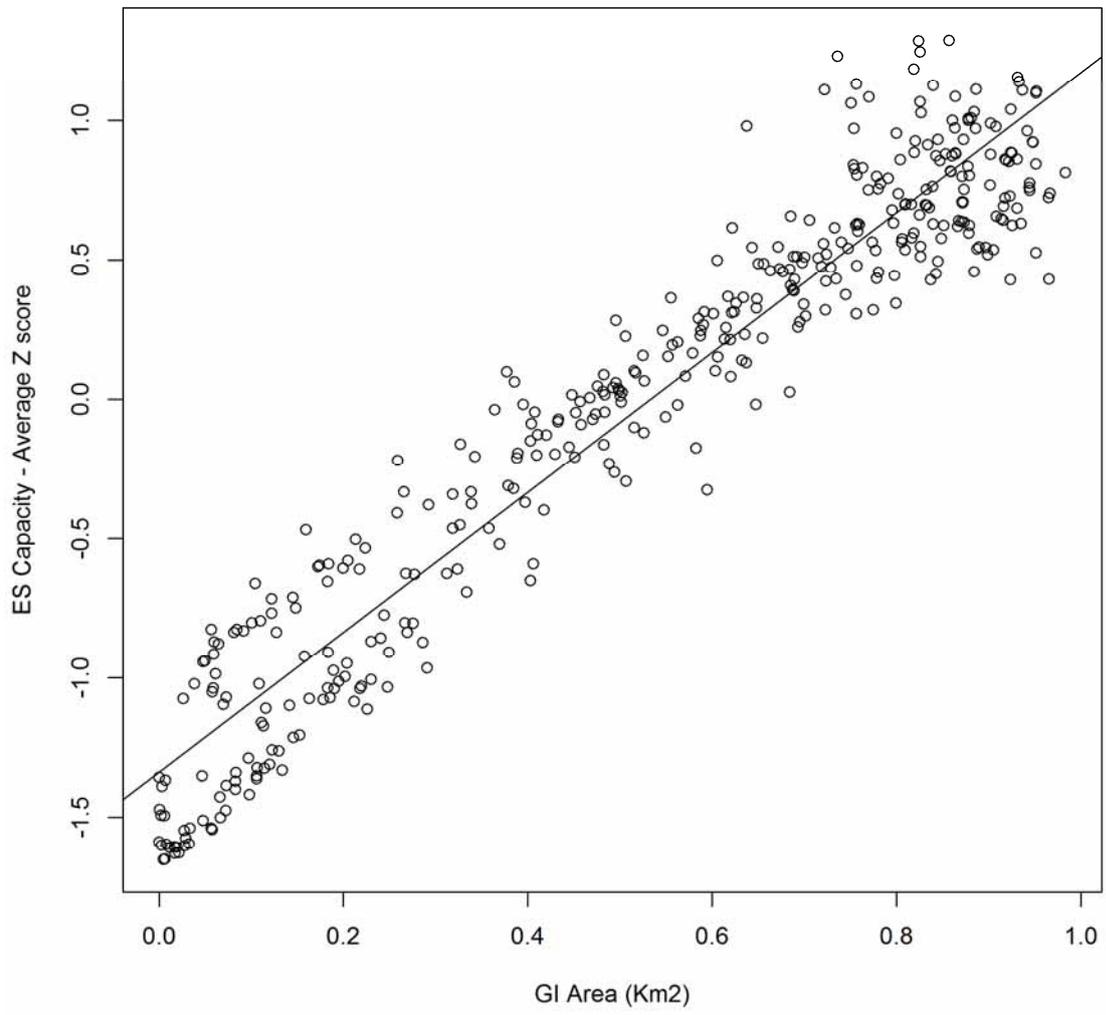

(b)

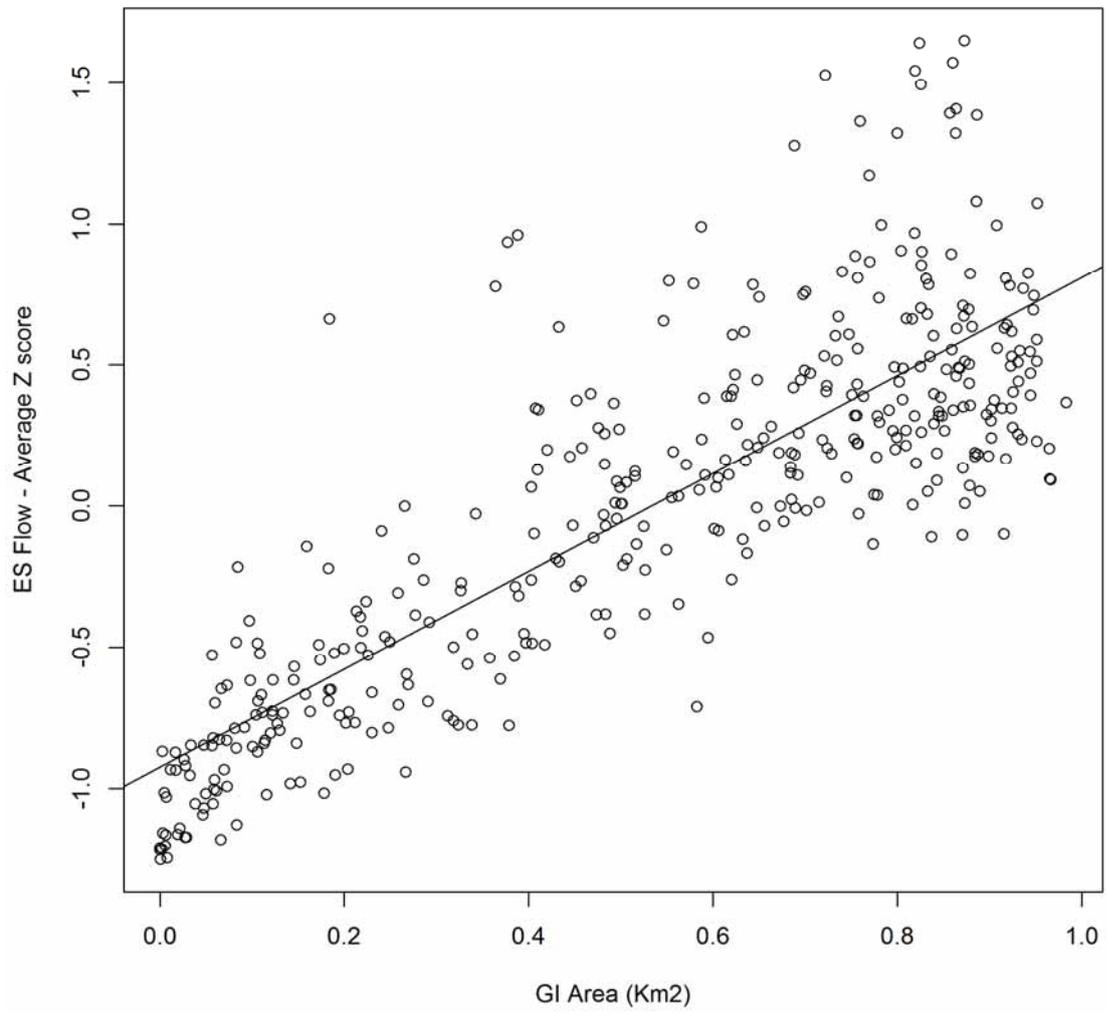

*Figure 8 – Regression analysis showing the multifunctionality of green infrastructure (GI) in terms of ES capacity ($R^2$=0.92; y = 2.51 x GI Area - 1.33; p < 0.0001) and flow ($R^2$ = 0.70; y = 1.73 x GI Area - 0.93; p < 0.0001).*

# 10. Supplementary Material

*Appendix A – Location of the study area within the Mediterranean region, and the developed Land use land cover (LULC) model.*

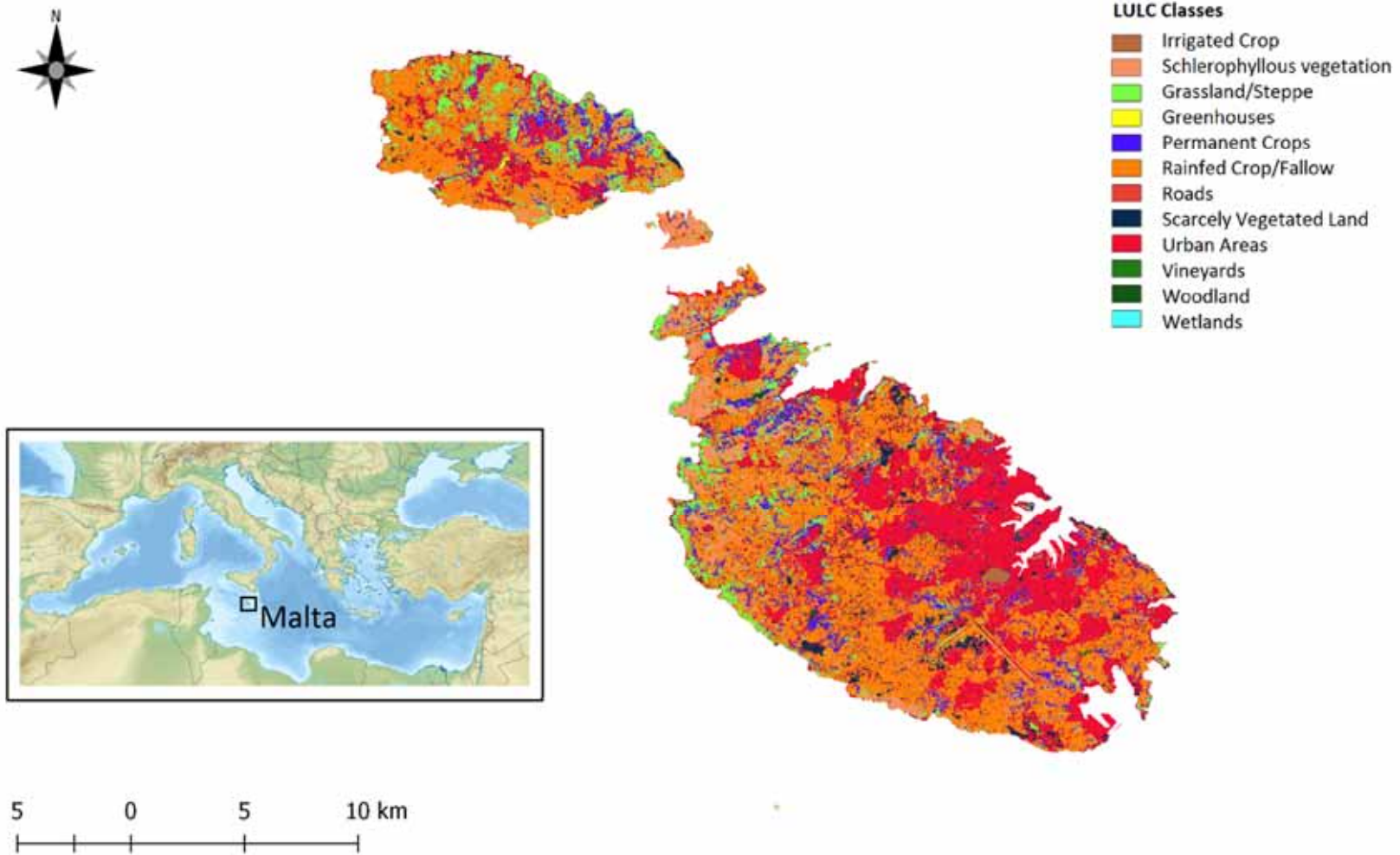

*Appendix B – The influence of environmental variables on $NO_2$ concentration in landscape buffers (r=500m). The difference in the deviance of the two compared models following backward elimination and significance using a $\chi^2$ test are shown. The column Effect shows the direction of the significant effects.*

| Independent variable | df | $\chi^2$ | p | Effect |
|---|---|---|---|---|
| Population | 1 | 8.32 | 0.004 | + |
| Primary Road Area | 1 | 9.27 | 0.002 | + |
| Residential Road Area | 1 | 27.53 | <0.0001 | + |
| Trunk Road Area | 1 | 13.56 | 0.0002 | + |
| Industrial Zones Area (Urban Atlas Class 12100) | 1 | 4.27 | 0.04 | + |
| Coastal | 1 | 12.29 | 0.0005 | + |
| Mean Elevation | 1 | 2.11 | 0.15 | - |
| Primary Road Area x Coastal | 1 | 5.55 | 0.02 | |
| Industrial Zones Area x Coastal | 1 | 5.57 | 0.02 | |
| Goodness of fit – $R^2$ (Nakagawa & Schielzeth, (2013)) | $R^2_m = 0.41$ $R^2_c = 0.75$ | | | |

*Appendix C – Environmental data with LULC categories area in a grid of 1km² cells was fitted over the PCA. The R² value is a measure of separation among the different levels of that variable, and its significance value was calculated using 1000 random permutations of the category levels.*

| LULC | PC1 | PC2 | $R^2$ | p |
|---|---|---|---|---|
| Irrigated arable land | 0.99733 | 0.07306 | 0.3428 | 0.001 |
| Garrigue | 0.38738 | -0.92192 | 0.2178 | 0.001 |
| Steppe | 0.6287 | -0.77765 | 0.1912 | 0.001 |
| Greenhouses | 0.9276 | 0.37357 | 0.0529 | 0.001 |
| Orchards and shrubs | 0.97719 | -0.21236 | 0.2025 | 0.001 |
| Non-irrigated arable land | 0.98735 | 0.15854 | 0.805 | 0.001 |
| Roads | 0.70255 | 0.71163 | 0.029 | 0.003 |
| Rocks | 0.79426 | 0.60758 | 0.0288 | 0.004 |
| Urban areas | -0.39631 | 0.91812 | 0.1377 | 0.001 |
| Vineyards | 0.99988 | 0.01535 | 0.1627 | 0.001 |
| Woodland | 0.40962 | -0.91225 | 0.019 | 0.02 |
| Wetlands | -0.13229 | -0.99121 | 0.011 | 0.106 |